\documentclass[12pt]{iopart}
\usepackage{iopams}  
\usepackage{color}
\usepackage{graphicx}
\usepackage{array}
\usepackage{braket}

\newcommand{\para}{{\mkern2mu\vphantom{\perp}\vrule depth 0pt\mkern2mu\vrule depth 0pt\mkern2mu}}

\newcommand{\xp}{\ensuremath{x_{\para}}}
\newcommand{\xs}{\ensuremath{x_{\perp}}}
\newcommand{\Lp}{\ensuremath{L_{\para}}}
\newcommand{\Ls}{\ensuremath{L_{\perp}}}
\newcommand{\Tc}{\ensuremath{T_{\mathrm{c}}}}
\newcommand{\Fres}{\ensuremath{F_{\mathrm{res}}}}

\newcommand{\Pf}{\ensuremath{\mathrm{Pf\,}}}

\newcommand{\ev}{\ensuremath{\mathrm{e}}}
\newcommand{\od}{\ensuremath{\mathrm{o}}}
\newcommand{\Bessel}[1]{\ensuremath{K_{#1}}}

\newcommand{\mystack}[2]{{\scriptstyle #1 \atop\scriptstyle #2}}

\newcommand{\FC}{\ensuremath{\mathcal{F}_\mathrm C}}

\begin{document}

\title[Critical Casimir force scaling functions of the two-dimensional Ising model]{Critical Casimir force scaling functions of the two-dimensional Ising model at finite aspect ratios}

\author{Hendrik Hobrecht \& Alfred Hucht}

\address{Universit{\"a}t Duisburg-Essen, Lotharstr.~1, D-47048 Duisburg}
\ead{fred@thp.uni-due.de}
\vspace{10pt}
\begin{indented}
\item[]\today
\end{indented}

\begin{abstract}
We present a systematic method to calculate the universal scaling functions for the critical Casimir force and the according potential of the two-dimensional Ising model with various boundary conditions.
Therefore we start with the dimer representation of the corresponding partition function $Z$ on an $L\times M$ square lattice, wrapped around a torus with aspect ratio $\rho=L/M$.
By assuming periodic boundary conditions and translational invariance in at least one direction, we systematically reduce the problem to a $2\times2$ transfer matrix representation.
For the torus we first reproduce the results by Kaufman and then give a detailed calculation of the scaling functions.
Afterwards we present the calculation for the cylinder with open boundary conditions.
All scaling functions are given in form of combinations of infinite products and integrals.
Our results reproduce the known scaling functions in the limit of thin films $\rho\to 0$.
Additionally, for the cylinder at criticality our results confirm the predictions from conformal field theory.
\end{abstract}

%
\vspace{2pc}

\section{Introduction}

The two-dimensional Ising model is one of the most popular and best understood systems in statistical physics.
Since Lars Onsager gave an exact solution for the bulk behavior of the infinite system \cite{Onsager44}, many more fascinating properties were found and calculated.
Its critical behavior has some remarkable features like scaling and universality \cite{Widom65,Kadanoff66}, and conformal invariance \cite{Polyakov70,Cardy84,CardyPeschel88}.
There are as many ways to calculate these properties as there were people working on this topic, nevertheless, in the following we want to focus on the dimer \textit{ansatz}, which uses the tight relation between the generating function of closest-packings of dimers on a lattice and the partition function of the Ising model, found by Kasteleyn \cite{Kasteleyn63} and exhaustively examined by McCoy \& Wu \cite{McCoyWu73}.
Their work gives an elegant and easy way to handle not only the thermodynamic limit but also finite systems and various boundary conditions, using the connection between the Pfaffian and the determinant of a sparse matrix.

In finite systems, the divergence of the bulk correlation length $\xi_\infty$ at criticality gives rise to the thermodynamic analogue of the quantum-electrodynamic Casimir effect, proposed by its eponym Hendrik B. G. Casimir in 1948 \cite{Casimir48}.
The thermodynamic Casimir effect was first introduced by Fisher \& de Gennes \cite{FisherdeGennes78}; they investigated the universal finite-size effects caused by critical fluctuations in a system which is confined in slab geometry.
Experimentally this effect was first proven to exist due to the measurement of the thickness of superfluid $^{4}$He films, which changes depending on the temperature close to the $\lambda$ transition and reveals a critical thinning of the film \cite{GarciaChan99}.
Ever since there were several other experimental realizations of systems showing thermodynamic Casimir forces, like tricritical $^{3}$He-$^{4}$He \cite{GarciaChan02} and binary liquids \cite{FukutoYanoPershan05}.
As a promising alternative to the film geometry, spherical objects in front of a surface \cite{HertleinHeldenGambassiDietrichBechinger08, SoykaZvyaHertHeldBech08} or many spherical objects in form of colloidal suspensions \cite{BonnOtwiSacaGuoWegSchall09,ZAB11,DVNBS13} were studied experimentally; those systems allow the direct measurement of the finite-size contribution to the free energy via the position distribution of the particles.

For a theoretical understanding, several Monte Carlo studies were made, using some kind of thermodynamic integration to investigate the finite-size scaling behavior \cite{Hucht07a,VasilyevGambassiMaciolekDietrich09}.
Later on, those Monte Carlo methods were generalized to spherical objects \cite{Hasenbusch13,Vasilyev14}, several using the same direct observation of the scaling behavior as the experiments \cite{HobrechtHucht14,HobrechtHucht15a}.

Additionally, at criticality itself, i.e., exactly at the critical temperature $T=\Tc$, conformal invariance as an extension of scale invariance arises and makes it possible to calculate the properties of systems having a vast variety of geometric constrains.
For $d$-dimensional systems this invariance is restricted to the group of M{\"o}bius transformations, which is sufficient to calculate at least the asymptotic behavior of two spherical objects, a spherical object in front of a wall, and similar configurations in a critical medium \cite{BurkhardtEisenriegler95,EisenrieglerRitschel95}, while for $d=2$ this group is infinite-dimensional due to complex analysis.
This plethora makes it possible to calculate the Casimir force between two arbitrary shaped objects \cite{BimonteEmigKardar13} and to investigate three-body interaction \cite{HobrechtHucht15a}.
To the best of our knowledge there is almost no work connecting the predictions from conformal field theory with the exact scaling functions at criticality for the two-dimensional Ising model except for a work on the M{\"o}bius strip and the Klein's bottle \cite{LuWu01}, albeit the first are known since long \cite{Philippe97}. 
In this context it is worth mentioning that Ferdinand and Fisher already calculated the according scaling functions at criticality on the torus in terms of elliptic $\vartheta$ functions in 1969 \cite{FerdinandFisher69,HuchtGruenebergSchmidt11}.

As mentioned above, we will focus on the two-dimensional Ising model on the torus and on the cylinder.
First we will recap the finite-size scaling regime and its connection to the thermodynamic Casimir effect, especially the direction dependency of the scaling variables.
Then we will start our calculation with the matrix derived by Kasteleyn for a system with $M$ rows of spin in parallel direction and $L$ columns of spins in perpendicular direction \cite{Kasteleyn63} and calculate its Pfaffian by means of its determinant.
Therefore we reduce the original $4LM\times4LM$ matrix to a $2\times2$ transfer matrix, assuming translational invariance and periodic boundary conditions in the parallel direction similar to the calculation in \cite{LuWu01}.
Afterwards we will derive first the partition function on the torus in a form similar to the one by Kaufman \cite{Kaufman49} and second the according universal critical Casimir potential and Casimir force scaling functions.
The latter calculation is an extension of the one already presented in \cite{HuchtGruenebergSchmidt11} to illuminate the details and especially the pitfalls of the necessary regularizations.
Next we will go back to the transfer matrix representation and focus on the partition function of the Ising model on the cylinder with open boundary conditions.
Eventually we will present the analogous calculation of the scaling functions and show how our results coincide in the limit $\rho\to0$ with the well-known scaling function of the thin strip with open boundary conditions \cite{EvansStecki94}.
Finally we compare the Casimir potential scaling function at criticality with the predictions of conformal field theory.

\section{The finite-size scaling regime}

Whenever a system with long-ranged thermodynamic fluctuations is confined by some geometrical constrains, its free energy $F$ becomes explicitly dependent on this geometry, characterized to leading order by the smallest length scale, e.g., there is a dependency on the thickness $\Ls$ of a $\Ls\times\Lp^{d-1}$ slab if $\Ls\ll\Lp$.
For such a system the reduced free energy $F(T;\Ls,\Lp)$, defined in units of $k_{\mathrm{B}}T$ with the Boltzmann constant $k_{\mathrm{B}}$, then may be decomposed as
\begin{equation}
	F(T;\Ls,\Lp)=F_{\infty}(T;\Ls,\Lp)+\Fres(T;\Ls,\Lp)
\end{equation}
into a part corresponding to the bulk and surface behavior of the infinite system $F_{\infty}(T;\Ls,\Lp)=Vf_{\mathrm{b}}(T)+Af_{\mathrm{s}}(T)$ with the surface area $A=\Lp^{d-1}$ and the volume $V=\Ls A$, where $f_{\mathrm{b}}$ is the bulk free energy per unit volume and $f_{\mathrm{s}}$ is the surface free energy per unit area, both depending only on the temperature $T$, and into a residual finite-size contribution $\Fres(T;\Ls,\Lp)$, which vanishes in the thermodynamic limit for $T\neq \Tc$.
The latter one is also called Casimir potential, especially in the context of colloidal suspensions, and both terms will be used synonymously in this work.
In the vicinity of the critical temperature $\Tc$ the bulk correlation length diverges as
\footnote{Throughout this work, the symbol $\simeq$ means ``asymptotically equal`` in the respective limit, $\Lp,\Ls\rightarrow\infty$, $T\rightarrow\Tc$, keeping the scaling variables $x$ and $\rho$ fixed, i.\,e., $f(L)\simeq g(L)\Leftrightarrow\lim_{L\rightarrow\infty}f(L)/g(L)=1$.}
\begin{equation}
	\xi_{\infty}(t)\stackrel{t>0}{\simeq}\xi_{+}t^{-\nu},
\end{equation}
where $t=T/\Tc-1$ is the reduced temperature and $\xi_{+}$ the correlation length amplitude above criticality, while $\nu$ is the according scaling exponent, with $\nu=1$ for the $2d$ Ising model.

Near the critical point and for arbitrary $\rho$, $\Fres$ only depends on the length ratios $\Ls/\xi_\infty(t)$ and $\Lp/\xi_\infty(t)$.
Following Fisher \& de Gennes \cite{FisherdeGennes78}, the residual free energy then fulfills a scaling \textit{ansatz} according to
\begin{equation}
	\Fres(T;\Ls,\Lp)\simeq \rho^{1-d} \Theta_{\perp}(\xs,\rho)
\end{equation}
with the Casimir potential scaling function $\Theta_{\perp}$, depending only on the temperature scaling variable and the aspect ratio 
\begin{equation}
\label{eq:xperp}
	\xs\equiv t\left(\frac{\Ls}{\xi_{+}}\right)^{\!\frac{1}{\nu}}
	\stackrel{t>0}{\simeq}\left(\frac{\Ls}{\xi_{\infty}(t)}\right)^{\frac{1}{\nu}},\qquad
	\rho\equiv\frac{\Ls}{\Lp}.
\end{equation}
If, on the other hand, $\Lp \lesssim \Ls$, the leading relevant quantity for finite-size effects is the ratio $\Lp/\xi_\infty$.
Hence we define the scaling function $\Theta_{\para}$ according to
\begin{equation}
	\Fres(T;\Ls,\Lp)\simeq \rho\,\Theta_{\para}(\xp,\rho),
\end{equation}
which depends on the corresponding scaling variable $\xp\equiv t(\Lp/\xi_+)^{1/\nu}$.
For a comparison with the results from conformal field theory we have to introduce a third quantity, namely the volume scaling function
\begin{equation}
\label{eq:FresVolume}
	\Fres(T;\Ls,\Lp)\simeq\Theta(x,\rho)
\end{equation}
with the generalized volume scaling variable 
\begin{equation}
	x\equiv t\left(\frac{V}{\xi^{d}_{+}}\right)^{\!\frac{1}{d\nu}}.
\end{equation}
Consequently, the three scaling functions fulfill the relations \cite{HuchtGruenebergSchmidt11}
\begin{equation}
\label{eq:ThetaPerpParaRelation}
	\Theta(x,\rho) = \rho \, \Theta_{\para}(\xp,\rho) = \rho^{1-d}\Theta_{\perp}(\xs,\rho).
\end{equation} 
Note that for the volume scaling function no direction is preferred, which intrinsically translates to the principles of conformal invariance.

The reduced critical Casimir force per area in perpendicular direction is given by the derivative
\begin{equation}
\FC(T;\Ls,\Lp) \equiv -\frac{1}{A}\frac{\partial}{\partial \Ls}\Fres(T;\Ls,\Lp)
\end{equation}
and near criticality scales like
\begin{equation}
\FC(T;\Ls,\Lp) 	\simeq \Ls^{-d} \vartheta_\perp(\xs, \rho) 
			\simeq \Lp^{-d} \vartheta_\para(\xp, \rho)
			\simeq V^{-1} \vartheta(x, \rho),
\end{equation}
defining the three Casimir force scaling functions $\vartheta_\perp$, $\vartheta_\para$ and $\vartheta$, suitable for the three cases $\rho\lesssim 1$, $\rho\gtrsim 1$ and $\rho\approx 1$, respectively.
The three Casimir force scaling functions fulfill the relations
\begin{equation}
\label{eq:thetaPerpParaRelation}
	\vartheta(x,\rho) = \rho \, \vartheta_{\para}(\xp,\rho) = \rho^{1-d}\vartheta_{\perp}(\xs,\rho)
\end{equation} 
analogous to (\ref{eq:ThetaPerpParaRelation}) and are related to the $\Theta$s via the scaling relations \cite{Dohm09,HuchtGruenebergSchmidt11}
\numparts
\begin{eqnarray}
\vartheta_{\perp}(\xs,\rho)&=&-\left[1-d+\frac{\xs}{\nu}\frac{\partial}{\partial \xs}+\rho\frac{\partial}{\partial\rho}\right]\Theta_{\perp}(\xs,\rho),\\
\label{eq:Thetathetapara}
\vartheta_{\para}(\xp,\rho)&=&-\left[1+\rho\frac{\partial}{\partial\rho}\right]\Theta_{\parallel}(\xp,\rho),\\
\vartheta               (x,\rho)&=&-\left[\frac{x}{d\nu}\frac{\partial}{\partial x}+\rho\frac{\partial}{\partial\rho}\right]\Theta(x,\rho).
\end{eqnarray}
\endnumparts

\section{From dimer to transfer matrix representation}

\begin{figure}[h]
	\centering
	\includegraphics[width=0.6\textwidth]{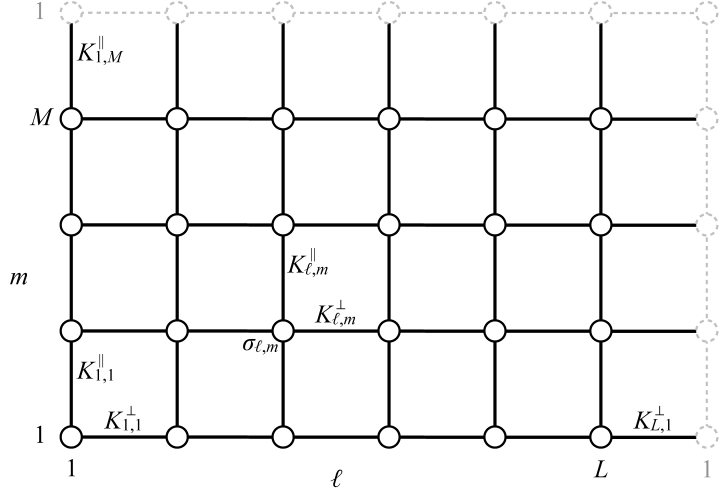}
	\caption{The square lattice with toroidal geometry for $M=4$ and $L=6$.}
	\label{fig:Lattice}
\end{figure}

From now on we will focus on the square lattice Ising model wrapped around a torus.
The partition function $Z=\rme^{-F}$ reads
\begin{equation}
\label{eq:PartitionHamiltonian}
	Z=\Tr\exp\sum_{\ell=1}^{L}\sum_{m=1}^{M}\left(K^{\perp}_{\ell,m}\sigma_{\ell,m}\sigma_{\ell+1,m}+K^{\para}_{\ell,m}\sigma_{\ell,m}\sigma_{\ell,m+1}\right),
\end{equation}
where $K^{\perp}_{\ell,m}$ and $K^{\para}_{\ell,m}$ are the reduced couplings between nearest neighbor spins in perpendicular and parallel direction, respectively.
For convenience we will use $L\equiv\Ls$ and $M\equiv\Lp$ for the according lengths of the system, see figure~\ref{fig:Lattice}, with $\sigma_{\ell,M+1}=\sigma_{\ell,1}$ and $\sigma_{L+1,m}=\sigma_{1,m}$ due to the periodic boundary conditions.

Within the dimer representation of this model, each spin translates to a vertex, which can be described by a $4\times4$ matrix \cite{McCoyWu73}.
Then the partition function reads
\begin{equation}
\label{eq:Z_with_Pfaffians}
	Z=\frac{1}{2}\left(\frac{2}{1-z^{2}}\right)^{LM}\left(\pm\Pf\mathbf{A}_{++}+\Pf\mathbf{A}_{+-}+\Pf\mathbf{A}_{-+}+\Pf\mathbf{A}_{--}\right),
\end{equation}
where the central quantity is the Pfaffian of the $4\times4$ antisymmetric block matrix
\begin{equation}
	\label{eq:BlockMatrixAab}
	\mathbf{A}_{ab}=\left[\begin{array}{cccc}
		\mathbf{0}						&\mathbf{1}+\tilde{\mathbf{U}}_{a}	&-\mathbf{1}						&-\mathbf{1}\\
		-\mathbf{1}-\tilde{\mathbf{U}}_{a}^{T}	&\mathbf{0}					&\mathbf{1}						&-\mathbf{1}\\
		\mathbf{1}						&-\mathbf{1}					&\mathbf{0}						&\mathbf{1}+\tilde{\mathbf{V}}_{b}\\
		\mathbf{1}						&\mathbf{1}					&-\mathbf{1}-\tilde{\mathbf{V}}_{b}^{T}	&\mathbf{0}
	\end{array}\right],
\end{equation}
with every entry being a $LM\times LM$ matrix, while $a,b \in \{+,0,-\}$ describe the boundary conditions in the two directions.
The matrices $\mathbf{\tilde{U}}_{a}=\mathbf{U} (\mathbf{H}_{a}\otimes\mathbf{1})$ and $\mathbf{\tilde{V}}_{b}=\mathbf{V} (\mathbf{1}\otimes\mathbf{H}_{b})$ combine the lattice structure represented by
\begin{equation}
	\mathbf{H}_{a}=
	\left(\begin{array}{ccccc}
		 0		&  1	& 0	& \cdots 	&  0 		\\
		 0		&  0 	& 1	& 		&  0 		\\
		 \vdots	&  	& 	& \ddots 	&  \vdots	\\
		 0		&  0 	& 0	&	 	&  1		\\
		 a		&  0	& 0	& \cdots 	&  0 		\\	
	\end{array}\right),
\end{equation}
and the couplings $\mathbf{U}=\mathbf{diag}(u_{1,1},u_{1,2},\dots,u_{L,M})$ and $\mathbf{V}=\mathbf{diag}(v_{1,1},v_{1,2},\dots,v_{L,M})$ in parallel and perpendicular direction, respectively, which are related to the reduced couplings in (\ref{eq:PartitionHamiltonian}) according to $u_{\ell,m}=\tanh K^{\para}_{\ell,m}$ and $v_{\ell,m}=\tanh K^{\perp}_{\ell,m}$.
The Pfaffian of a matrix can be calculated using the relation 
\begin{equation}
	\Pf\mathbf{A}_{ab}=\pm\sqrt{\det\mathbf{A}_{ab}},
\end{equation}
where the sign in the partition function can be determined by continuation from $T=0$.

Assuming that the parallel direction is not only periodic but also translationally invariant, i.e., both $u_{\ell,m}=u_{\ell}$ and $v_{\ell,m}=v_{\ell}$ depend only on the first index, we can reduce the calculation of the determinant of the matrix $\mathbf{A}_{ab}$ to the product
\begin{equation}
	\det\mathbf{A}_{ab}=\prod_{m=1}^{M}\det\mathbf{B}_{b}(\varphi^{a}_{m}),
\end{equation}
with the $4L\times 4L$ block matrices
\begin{equation}
\label{eq:DefBbphi}
	\mathbf{B}_{b}(\varphi)=\left[\begin{array}{cccc}
		\mathbf{0}				& \mathbf{S}_{+}(\varphi)	& -\mathbf{1}		& -\mathbf{1}\\
		-\mathbf{S}_{-}(\varphi)	& \mathbf{0}			& \mathbf{1}		& -\mathbf{1}\\
		\mathbf{1}				& -\mathbf{1}			& \mathbf{0}		& \mathbf{J}_{b} \\
		\mathbf{1}				& \mathbf{1}			& -\mathbf{J}_{b}^{T}	& \mathbf{0}
	\end{array}\right],
\end{equation}
using the eigenvalues $\varphi^{a}_{m}$ of $\mathbf{H}_{a}$, with $\varphi^{+}_{m}=2m\pi/M$ and $\varphi^{-}_{m}=(2m-1)\pi/M$, running over even and odd integers, respectively, to bring $\mathbf{A}_{ab}$ into block diagonal form.
Due to the translational invariance the matrices $\mathbf{U}$ and $\mathbf{V}$ reduce to $\mathbf{u}=\mathbf{diag}(u_{1},\dots,u_{L})$ and $\mathbf{v}=\mathbf{diag}(v_{1},\dots,v_{L})$, respectively.
Here $\mathbf{S}_{\pm}(\varphi)=\mathbf{1}+\mathbf{u}\,\rme^{\pm\rmi\varphi}$ represents the structure and couplings in parallel direction, while the matrix $\mathbf{J}_{b}=\mathbf{1}+\mathbf{v} \mathbf{H}_{b}$ is basically the reduced version of the block matrix element $\mathbf 1 + \mathbf{\tilde{V}}_b$ in (\ref{eq:BlockMatrixAab}).
In the following we will drop the $\varphi$ dependence to improve readability.

The block matrices (\ref{eq:DefBbphi}) can be further simplified using a Schur reduction
\begin{equation}
\label{eq:Schur2}
	\det\left[\begin{array}{cc}
		\mathbf{a}_{11} & \mathbf{a}_{12} \\ \mathbf{a}_{21} & \mathbf{a}_{22}
	\end{array}\right] = \det\mathbf{a}_{11}-\det(\mathbf{a}_{22}-\mathbf{a}_{21}\mathbf{a}_{11}^{-1}\mathbf{a}_{12}),
\end{equation}
leading to
\begin{equation}
	\det\mathbf{B}_{b}=\det\left[\begin{array}{cc}\mathbf{0}&\mathbf{S}_{+}\\-\mathbf{S}_{-}&\mathbf{0}\end{array}\right]\det\left[\begin{array}{cc}-\mathbf{\Delta}&\mathbf{J}_{b}-\mathbf{\Sigma}\\\mathbf{\Sigma}-\mathbf{J}_{b}^{T}&\mathbf{\Delta}\end{array}\right],
\end{equation}
with $\mathbf{\Sigma}\equiv\mathbf{S}_{+}^{-1}+\mathbf{S}_{-}^{-1}$ and $\mathbf{\Delta}\equiv\mathbf{S}_{+}^{-1}-\mathbf{S}_{-}^{-1}$.
As the matrices $\mathbf{S}_{+}$ and $\mathbf{S}_{-}$ are diagonal, the first determinant is simply
\begin{equation}
	\det\left[\begin{array}{cc}\mathbf{0}&\mathbf{S}_{+}\\-\mathbf{S}_{-}&\mathbf{0}\end{array}\right]=\prod_{\ell=1}^{L}\left(1+u_{\ell}\rme^{\rmi\varphi}\right)\left(1+u_{\ell}\rme^{-\rmi\varphi}\right).
\end{equation}
With a second Schur reduction, the second determinant simplifies to
\begin{equation}
	\det\left[\begin{array}{cc}-\mathbf{\Delta}&\mathbf{J}_{b}-\mathbf{\Sigma}\\\mathbf{\Sigma}-\mathbf{J}_{b}^{T}&\mathbf{\Delta}\end{array}\right]
	=\det(-\mathbf{\Delta})\det \mathbf{C}_b,
\end{equation}
with Schur complement
\begin{equation}
\mathbf{C}_b = \mathbf{\Delta}+\left(\mathbf{\Sigma}-\mathbf{J}_{b}^{T}\right)\mathbf{\Delta}^{-1} \left(\mathbf{J}_{b}-\mathbf{\Sigma}\right).
\end{equation}
Since $\mathbf{\Delta}$ is diagonal, its determinant is
\begin{equation}
	\det(-\mathbf{\Delta})=\prod_{\ell=1}^{L}\left(\frac{1}{1+u_{\ell}\rme^{-\rmi\varphi}}-\frac{1}{1+u_{\ell}\rme^{\rmi\varphi}}\right),
\end{equation}
and we finally get
\begin{equation}
	\det\mathbf{B}_{b}=(2\rmi\sin\varphi)^{L} \det \mathbf{u} \det \mathbf{C}_b.
\end{equation}

The factor $(2\rmi\sin\varphi)^{L}$ can be included in $\tilde\mathbf C_b = 2\rmi\sin\varphi \, \mathbf C_b$, leading to the symmetric, cyclic, tridiagonal matrix 
\begin{eqnarray}
\label{eq:TridiagonalC}
	\tilde\mathbf{C}_{b} =
	\left(\begin{array}{ccccc}
		a_{1}	& b_{1} 	&	 	& b_{L}\\
		b_{1}	& a_{2} 	& \ddots	& \\
				& \ddots	& \ddots	& b_{L-1}\\
		b_{L}	&		& b_{L-1}	& a_{L}
	\end{array}\right),
\end{eqnarray}
which can be calculated  using a $2\times2$ transfer matrix $\mathbf{\tilde{T}}_{\ell}$ according to
\begin{equation}
\label{eq:TridiagonalToTM}
	\det\tilde\mathbf{C}_{b}=2(-1)^{L+1}\prod_{\ell=1}^{L}b_{\ell}+\Tr(\tilde{\mathbf{T}}_{L}\tilde{\mathbf{T}}_{L-1}\cdots\tilde{\mathbf{T}}_{2}\tilde{\mathbf{T}}_{1}).
\end{equation}
The matrix elements of $\tilde\mathbf{C}_{b}$ are
\numparts
\begin{eqnarray}
	a_{\ell}=(q_{\ell-1}v_{\ell-1})^{2}\mu^{+}_{\ell-1}(\varphi)-\mu^{-}_{\ell}(\varphi)\\
	b_{\ell}=-q_{\ell}v_{\ell}(u_{\ell}^{-1}-u_{\ell}),
\end{eqnarray}
\endnumparts
with the functions $\mu^{\pm}_{\ell}(\varphi)=2\cos\varphi\pm(u_{\ell}^{-1}+u_{\ell})$, while
 $q_{L}=b$ and $q_{\ell\neq L}=1$ distinguishes between the boundary contributions and the rest of the system.
The transfer matrix then has the form
\begin{equation}
	\mathbf{\tilde{T}}_{\ell}=\left(\begin{array}{cc}
		a_{\ell}	& -b_{\ell-1}^{2} \\
		1	& 0
	\end{array}\right),
\end{equation}
mixing up couplings of different directions and different rows.
Nevertheless, it can be decomposed into three parts, separating the three couplings $u_{\ell}$, $v_{\ell-1}$ and $u_{\ell-1}$, as
\begin{equation}
	\mathbf{\tilde{T}}_{\ell}=\left(\begin{array}{cc}
	-\mu^{-}_{\ell} & 1 \\ 1 & 0
	\end{array}\right) \left(\begin{array}{cc}
	1 & 0 \\	 0	& (q_{\ell-1}v_{\ell-1})^{2}
	\end{array}\right) \left(\begin{array}{cc}
	1 & 0 \\ \mu^{+}_{\ell-1}	& -(u_{\ell-1}^{-1}-u_{\ell-1})
	\end{array}\right).
\end{equation}
Now we rearrange the three matrices into two matrices representing only one single coupling by combining the first and the third matrix for the same index $\ell$ to
\begin{equation}
	\mathbf{T}^{\para}_{\ell}=\left(\begin{array}{cc}
	-\mu^{-}_{\ell}(\varphi)	& 1 \\
	4\sin^{2}(\varphi)	& \mu^{+}_{\ell}(\varphi)
	\end{array}\right) \qquad
	\mathbf{T}^{\perp}_{\ell}=\left(\begin{array}{cc}
	1	& 0 \\
	0	& (q_{\ell}v_{\ell})^{2}
	\end{array}\right).
\end{equation}
Thus we finally get
\begin{equation}
\label{eq:detBbTorusFinal}
	\det\mathbf{B}_{b}(\varphi) = \det \mathbf{u} 
	\,\Big[\Tr\left(\mathbf{T}^\perp_L \mathbf{T}^\para_L \cdots \mathbf{T}^\perp_1 \mathbf{T}^\para_1\right)
	-2b\prod_{\ell=1}^{L}v_{\ell}(u_{\ell}^{-1}-u_{\ell})\Big].
\end{equation}

For the homogeneous and anisotropic case, i.e., $u_{\ell}=u$ and $v_{\ell}=v$ for all $\ell$, the transfer matrix can be symmetrized according to
\begin{equation}
	\mathbf{\bar{T}}=\sqrt{\mathbf{T}^{\perp}} \mathbf{T}^{\para} \sqrt{\mathbf{T}^{\perp}}=\mathbf{X} \left(\begin{array}{cc} \lambda^{+} & 0 \\ 0 & \lambda^{-} \end{array}\right) \mathbf{X}^{-1}
\end{equation}
with the unitary matrix
\begin{equation}
	\mathbf{X}=\left(\begin{array}{cc} \frac{\lambda^{+}-v^{2}\mu^{+}}{4v\sin^{2}(\varphi)} & \frac{\lambda^{-}-v^{2}\mu^{+}}{4v\sin^{2}(\varphi)} \\ 1 & 1 \end{array}\right),
\end{equation}
where $\lambda^{\pm}$ are the eigenvalues of $\mathbf{\bar{T}}$.
If additionally the system is isotropic, i.e., $u_\ell=v_\ell=z$, those eigenvalues are given by
\begin{equation}
\label{eq:TransferEigenvalues}
	\lambda^{\pm}(\gamma)=(1-z^{2})\rme^{\pm\gamma},
\end{equation}
where we have introduced $\gamma$, which already occurred in Onsager's famous solution of the Ising model on the torus \cite{Onsager44}, and is connected to the eigenvalues $\varphi$ by the relation
\begin{equation}
\label{eq:gammarelation}
	\cosh\gamma+\cos\varphi=\frac{(1+z^{2})^{2}}{2z(1-z^{2})}.
\end{equation}

\section{Scaling functions on the torus}

For an Ising system with homogeneous and isotropic couplings $z=\tanh K$ on the torus the determinant from the last chapter is given by
\begin{equation}
	\det\mathbf{A}_{ab}=z^{LM}(1-z^{2})^{LM}\prod_{m=1}^{M}2\left[\cosh(L\gamma^{a}_{m})-b\right].
\end{equation}
Following the discussion by McCoy \& Wu on the signs of the Pfaffians, the partition function of this geometry is given by
\begin{eqnarray}
	Z&=\frac{1}{2}\left(\frac{2}{1-z^{2}}\right)^{LM}\nonumber\\
	&\times\left[\pm\sqrt{\det\mathbf{A}_{++}}+\sqrt{\det\mathbf{A}_{+-}}+\sqrt{\det\mathbf{A}_{-+}}+\sqrt{\det\mathbf{A}_{--}}\right],
\end{eqnarray}
where the $+/-$ sign is used below/above the critical point.
For convenience we introduce the shorter form
\begin{equation}
	Z^{\pm}_{\ev/\od}=\prod_\mystack{0\,\leq\,m\,<\,2M}{m \,\mathrm{even/odd}} \left(\rme^{\frac{L}{2}\gamma_{m}}\pm\rme^{-\frac{L}{2}\gamma_{m}}\right),
\end{equation}
where we have used the symmetry $\gamma_{0}=\gamma_{2m}$.
Now the partition function $Z$ can be written as
\begin{equation}
\label{eq:PartitionFunctionTorus}
	Z=\frac{1}{2}\left(\frac{4z}{1-z^{2}}\right)^{LM/2}\left(Z^{+}_{\od} + Z^{-}_{\od} + Z^{+}_{\ev} \pm Z^{-}_{\ev}\right),
\end{equation}
which is just the solution by Kaufman \cite{Kaufman49} and was the starting point for the calculation in \cite{HuchtGruenebergSchmidt11}.
In the following the latter one will be illuminated in more detail.

We first note that in every $Z^{\pm}_{\ev/\od}$ there is either an even or an odd contribution 
\begin{equation}
	\label{eq:Q}
	Q_{\ev/\od}(L,M)=\prod_\mystack{0\,<\,m\,<\,2M}{m \,\mathrm{even/odd}}\rme^{\frac{L}{2}\gamma_{m}}
\end{equation}
that diverges exponentially fast with growing $L$.
To find the relevant contribution to the Casimir potential $\Fres$ we separate (\ref{eq:Q}) from the remaining parts and get
\numparts
\begin{eqnarray}
\label{eq:Zpmdelta}
	Z^{\pm}_{\ev}&=P^{\pm}_{\ev}(L,M)\left(\rme^{\frac{L}{2}\gamma_{0}}\pm\rme^{-\frac{L}{2}\gamma_{0}}\right)Q_{\ev}(L,M),\\
	Z^{\pm}_{\od}&=P^{\pm}_{\od}(L,M)Q_{\od}(L,M),
\end{eqnarray}
\endnumparts
where we follow the notation in \cite{FerdinandFisher69} for
\begin{equation}
	P^{\pm}_{\ev/\od}(L,M)=\prod_\mystack{0\,<\,m\,<\,2M}{m \,\mathrm{even/odd}}\left(1\pm\rme^{-L\gamma_{m}}\right).
\end{equation}
Now we split off the term $Q_{\od}(L,M)$, from (\ref{eq:PartitionFunctionTorus}) which is slightly larger than $Q_{\ev}(L,M)$, and get 
\begin{eqnarray}
	Z^{+}_{\od} + Z^{-}_{\od} + Z^{+}_{\ev} \pm Z^{-}_{\ev}\nonumber\\
	\quad =Q_{\od}\left\{P^{+}_{\od} + P^{-}_{\od} + \Delta Q\left[(\rme^{\frac{L}{2}\gamma_{0}}+\rme^{-\frac{L}{2}\gamma_{0}})P^{+}_{\ev} \pm (\rme^{\frac{L}{2}\gamma_{0}}-\rme^{-\frac{L}{2}\gamma_{0}})P^{-}_{\ev}\right]\right\},
\end{eqnarray}
with the quotient of the even and odd terms $\Delta Q(L,M)=Q_{\ev}(L,M)/Q_{\od}(L,M)$.

The leading terms of $\Delta Q(L,M)$ for $L\to\infty$, $M\to\infty$ were already calculated by Ferdinand and Fisher \cite{FerdinandFisher69} up to $\Or(\xp^{2})$ as
\begin{equation}
\label{eq:DeltaQApprox}
	\ln\Delta Q(L,M)\simeq -\rho\left[\frac{\pi}{4}+\xp^{2}\frac{\ln 2}{2\pi}+\Or(\xp^{4})\right],
\end{equation}
but for the scaling form we need an explicit form of the higher-order terms.
This contribution to the partition function is exponentially small, since $Q_{\od}>Q_{\ev}$, as noted before.
As we have changed to the logarithm, the product transforms into a sum, where the quotient translates into an alternating one, which can be recasted into
\begin{equation}
	\label{eq:Qdefinition}
	\ln\Delta Q(L,M)=\frac{L}{2}\left(2\sum_{m=1}^{M-1}\gamma_{2m}-\sum_{m=1}^{2M-1}\gamma_{m}\right).
\end{equation}
If we go to the scaling limit $L\to\infty$, $M\to\infty$ with $L/M=\rho$ we get
\begin{equation}
	\label{eq:gammascaling}
	\lim_\mystack{L,M\to\infty}{L/M=\rho} L\gamma_{m} = \rho\sqrt{\xp^{2}+\pi^2 m^2},
\end{equation}
with $\xp\equiv (1-z/z_{\mathrm{c}})M/\xi^{z}_{+}$, $z_{\mathrm{c}}=\sqrt{2}-1$, and $\xi^{z}_{+}=\frac{1}{2}$ \cite{McCoyWu73}.
To go to this limit we have to use the symmetry of $\gamma$ around $m=M$, so that the sums run over monotonic increasing functions.
Nevertheless, with (\ref{eq:gammascaling}) each of the two sums would be divergent and thus we have to subtract the leading divergent terms in $m$ as $M\to\infty$ and find the regularized form to be
\begin{equation}
	\gamma_{m}\sim\frac{1}{M}\left(\sqrt{x^{2}+\pi^{2}m^{2}}-\pi |m|-\frac{x^{2}}{2\pi |m|}\right).
\end{equation}
In fact one can easily see that this regularization leaves only the terms $\Or(\xp^{4})$, so we use it together with (\ref{eq:DeltaQApprox}) to calculate the scaling form
\begin{eqnarray}
	\fl\ln \Delta Q(\xp,\rho)\equiv-\rho\left[\frac{\pi}{4}+\xp^{2}\frac{\ln 2}{2\pi}-\sum_{m=1}^{\infty}\left(2\sqrt{\xp^{2}+4\pi^{2}m^{2}}-\sqrt{\xp^{2}+\pi^{2}m^{2}}-3\pi |m|\right)\right].
\end{eqnarray}
The application of the Poisson sum formula for the sum introduces the modified Bessel function of first kind $\Bessel{\nu}(x)$,
\numparts
\begin{eqnarray}
	\sum_{m=1}^{\infty}\left(2\sqrt{\xp^{2}+4\pi^{2}m^{2}}-\sqrt{\xp^{2}+\pi^{2}m^{2}}-3\pi |m|\right)\\
	\quad=\sum_{k=1}^{\infty}\left[\frac{3}{2\pi k^{2}}+\frac{|\xp|}{\pi k}\Bessel{1}\left(2 k |\xp|\right)-\frac{2|\xp|}{\pi k}\Bessel{1}\left(k |\xp|\right)\right]+\xp^{2}\frac{\ln 2}{2\pi}-\frac{|\xp|}{2}\\
	\quad=\frac{1}{\pi}\sum_{k=1}^{\infty}\frac{|\xp|}{k}\left[\Bessel{1}(2k|\xp|)-2\Bessel{1}(k|\xp|)\right]+\frac{\pi}{4}+\xp^{2}\frac{\ln 2}{2\pi}-\frac{|\xp|}{2}.
\end{eqnarray}
\endnumparts
Now we use the integral representation of $\Bessel{1}(\alpha k|\xp|)$ \cite[eq.~(3.389-4)]{GradshteynRyzhik80},
\begin{equation}
	\frac{|\xp|}{k}\Bessel{1}(\alpha k |\xp|)=\frac{1}{k}\int\limits_{0}^{\infty}\mathrm{d}t\,\mathrm{e}^{-\alpha k\sqrt{t^{2}+\xp^{2}}},
\end{equation}
and exchange the sum and the integral for the summation over $k$, which gives
\begin{equation}
	\sum_{k=1}^{\infty}\frac{1}{k}\int\limits_{0}^{\infty}\mathrm{d}t\,\mathrm{e}^{-\alpha k\sqrt{t^{2}+\xp^{2}}}
	=-\int\limits_{0}^{\infty}\mathrm{d}t\,\ln\left(1-\mathrm{e}^{-\alpha \sqrt{t^{2}+\xp^{2}}}\right).
\end{equation}
At this point it is convenient to introduce the integrals
\begin{equation}
	I_{\pm}(\xp)\equiv\frac{1}{2}\int\limits_{-\infty}^{\infty}\rmd\omega\,\ln\left(1\pm\rme^{-\sqrt{\xp^{2}+\pi^{2}\omega^{2}}}\right)
\end{equation}
with which we conclude with
\begin{equation}
\label{eq:ScalingQ}
	\ln\Delta Q(\xp,\rho)=-\rho\left[\frac{|\xp|}{2}+I_{+}(\xp)-I_{-}(\xp)\right].
\end{equation}
for the scaling limit of $\ln\Delta Q(L,M)$.

\begin{figure}[t]
	\centering
	\includegraphics[width=0.9\textwidth]{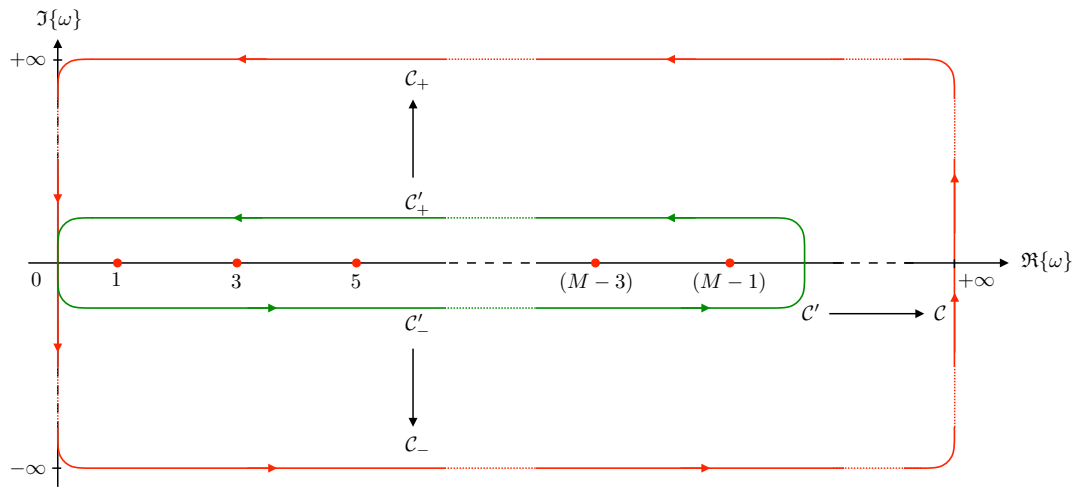}
	\caption{Contours used for the integrals in the finite case (\ref{eq:Qoddfinite}) and in the scaling regime (\ref{eq:limQodd}).
	The inner contour (green) $\mathcal{C}'$ can be decomposed into $\mathcal{C}_{+}'$ and $\mathcal{C}_{-}'$ in the upper and lower half-plane, respectively.
	For the calculation of the scaling limit the contour is deformed in such a way that $\Im\{\omega\}\to\pm\infty$ and $\Re\{\omega\}\to+\infty$.
	This outer (red) contour $\mathcal{C}$ can be analogously decomposed into $\mathcal{C}_{+}$ and $\mathcal{C}_{-}$.
	}
	\label{fig:ContourDeforming}
\end{figure}

Now we will turn to the calculation of the finite-size contribution of $\ln Q_{\od}$ to the Casimir potential $\Fres$ and its scaling form.
Therefore we again utilize the symmetry of $\gamma_{m}$ and recast the sum established by the logarithm to be of a monotonic increasing function in $m$.
For the scaling limit we need to expand the sum to infinity, which is easier to do in an integral representation.
Thus we first rewrite the sum over the odd integers into the difference of a sum and its alternating counterpart 
\begin{equation}
	\ln Q_{\od}(L,M)=\frac{L}{2}\left[\sum_{m=1}^{M-1}\gamma_{m}-\sum_{m=1}^{M-1}(-1)^{m}\gamma_{m}\right],
\end{equation}
which each can be written as a contour integral:
For the ordinary sum we have to integrate over $\pi\cot(\pi m)\gamma_{m}$ and for the alternating one over $\pi\csc(\pi m)\gamma_{m}$, together leading to 
\begin{equation}
\label{eq:Qoddfinite}
	\ln Q_{\od}(L,M)=-\frac{L}{2\pi\rmi}\oint\limits_{\mathcal{C}'}\rmd\omega\,\frac{\pi}{2}\tan\left(\frac{\pi\omega}{2}\right)\gamma_{\omega}
\end{equation}
with $\mathcal{C}'$ being the counterclockwise contour around the integers in the interval $[1,M-1]$.
As we already used the symmetry of $\gamma$, we can go to the scaling limit by deforming the contour $\mathcal{C}'$ to $\mathcal{C}$ as depicted in figure~\ref{fig:ContourDeforming}.
Thus we find 
\begin{equation}
	\label{eq:limQodd}
	\lim_\mystack{L,M\to\infty}{L/M=\rho} \ln Q_{\od}(L,M) = -\frac{\rho}{4\rmi}\oint\limits_{\mathcal{C}}\rmd\omega\tan\left(\frac{\pi\omega}{2}\right)\sqrt{\xp^{2}+\pi^{2}\omega^{2}}.
\end{equation}
Along the upper and lower part of the contour we find $\tan(\pi\omega/2)\stackrel{\omega\to\pm\rmi\infty}{\simeq}\pm\rmi$ and so the integral diverges, but this part just coincides with the integral representation of the thermodynamic limit, and so we use it as regularization to obtain the finite-size part $\ln\delta Q_{\od}$.
Subsequently we notice that the integral along the upper contour $\mathcal{C}_{+}$ and along the lower contour $\mathcal{C}_{-}$ are the complex conjugate of each other, while the integral along the line at $\Re\{\omega\}\to\infty$ and, due to the regularization, along the line at $\Im\{\omega\}\to\pm\infty$ all vanish.
Then the scaling form of $\delta Q_{\od}$, i.e., the regularized form of (\ref{eq:limQodd}), is only twice the real part of integral along the positiv imaginary axis
\begin{equation}
	\ln\delta Q_{\od}(\xp)=\Re\left\{-\frac{\rho}{2\rmi}\int\limits_{0}^{\rmi\infty}\rmd\omega\,\left[\tan\left(\frac{\pi\omega}{2}\right)-\rmi\right]\sqrt{\xp^{2}+\pi^{2}\omega^{2}}\right\},
\end{equation}
which, after some algebra, has the simple form
\begin{equation}
	\label{eq:IntegralPlus}
	\ln\delta Q_{\od}(\xp)=\rho I_{+}(\xp).
\end{equation}

For the scaling form of $P^{\pm}_{\delta}$ we follow the steps in \cite{FerdinandFisher69} and use the series expansion of the logarithm to find
\begin{equation}
	\ln P^{\pm}_{\ev/\od}(L,M)=-\sum_{n=1}^{\infty}\frac{(\mp1)^{n}}{n}\sum_\mystack{0\,<\,m\,<\,2M}{m \,\mathrm{even/odd}} \rme^{-L n \gamma_{m}}.
\end{equation}
For the scaling limit we have to use the symmetry of $\gamma$ again, which, assuming $M$ is even, gives an exponentially small and thus vanishing error in $\ln P^{\pm}_{\ev}$, and we get 
\numparts
\begin{eqnarray}
	\lim_\mystack{L,M\to\infty}{L/M=\rho}\ln P^{\pm}_{\ev/\od}(L,M)
	&=-2\sum_{n=1}^{\infty}\frac{(\mp1)^{n}}{n}\sum_\mystack{0\,<\,m\,<\,\infty}{m \,\mathrm{even/odd}}\rme^{-n\rho\sqrt{\xp^{2}+\pi^{2}m^{2}}}\\
	\label{eq:ScalingPpmdeltaSquare}
	&=2\sum_\mystack{0\,<\,m\,<\,\infty}{m \,\mathrm{even/odd}}\ln\left(1\pm\rme^{-\rho\sqrt{\xp^{2}+\pi^{2}m^{2}}}\right).
\end{eqnarray}
\endnumparts
Combining the term for the even indices in (\ref{eq:Zpmdelta}) with (\ref{eq:ScalingPpmdeltaSquare}) and changing from the square to a symmetric product from $-\infty$ to $\infty$ finally gives the scaling form of $P^{\pm}_{\ev/\od}$ as
\begin{equation}
	\label{eq:ScalingPpmdelta}
	P^{\pm}_{\ev/\od}(\xp,\rho)\equiv\prod_\mystack{m=-\infty}{m \,\mathrm{even/odd}}^{\infty}\!\!\left(1\pm\rme^{-\rho\sqrt{\xp^{2}+\pi^{2}m^{2}}}\right)
\end{equation}

\begin{figure}
	\centering
	\hfill%
	\includegraphics[width=0.45\textwidth]{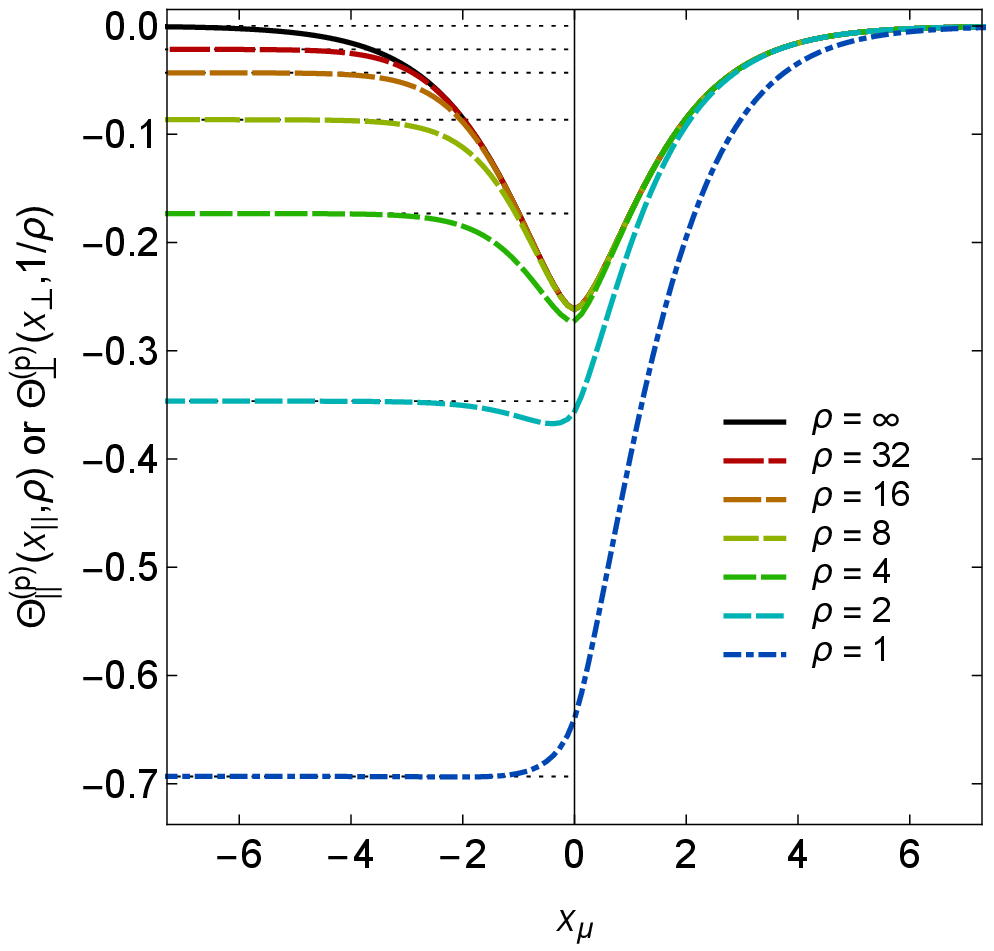}%
	\hfill\hfill%
	\includegraphics[width=0.45\textwidth]{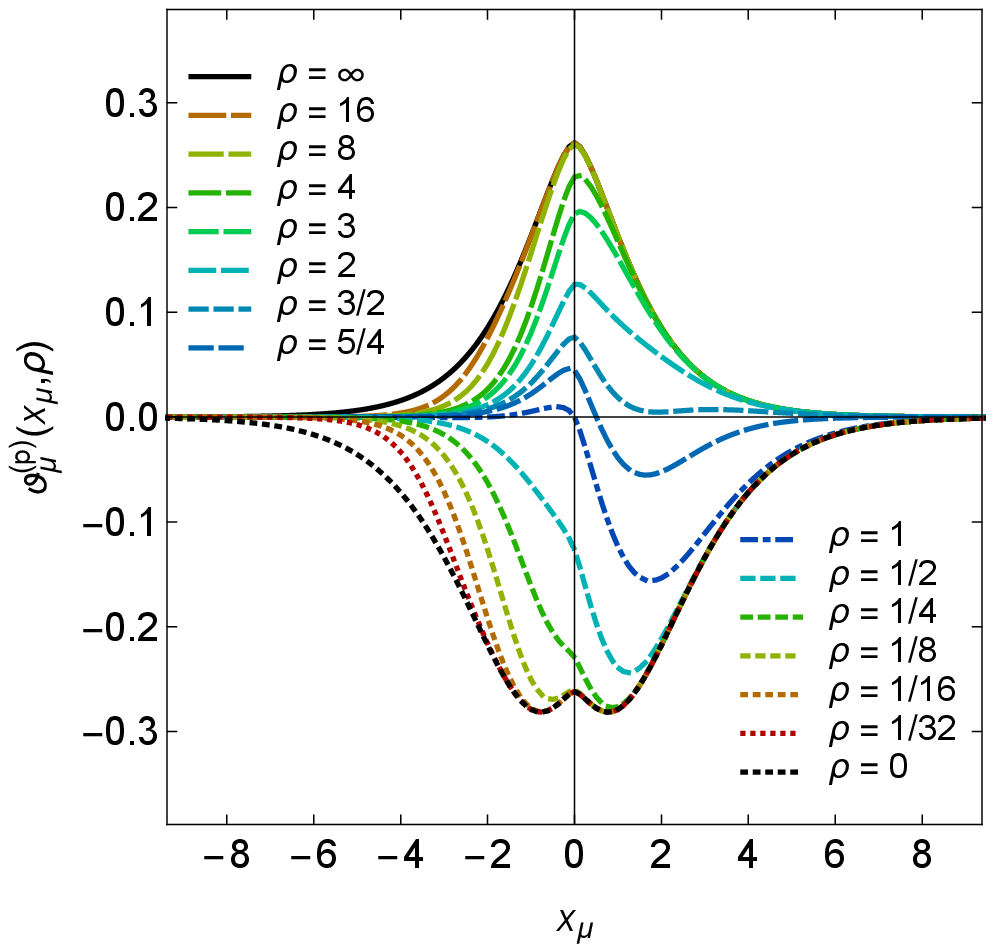}%
	\hfill{}
	\caption{Left: Casimir potential scaling function $\Theta^\mathrm{(p)}_\mu(x_\mu,\rho)$.
	Note that due to the periodic boundary conditions there is a symmetry $\Theta^{\mathrm{(p)}}_{\para}(\xp)=\Theta^{\mathrm{(p)}}_{\para}(\xs,1/\rho)$.
	Right: Casimir force scaling function $\vartheta^\mathrm{(p)}_\mu(x_\mu,\rho)$.
	We show $\vartheta_{\para}(\xp)$ for $\rho>1$ and $\vartheta_{\perp}(\xp)$ for $\rho<1$.
	}
\end{figure}

With (\ref{eq:IntegralPlus}), (\ref{eq:ScalingQ}) and (\ref{eq:ScalingPpmdelta}) we conclude with the scaling function of the free energy as a function of the scaling variable $\xp$ in parallel direction
\begin{equation}
\label{eq:Thetap}\fl\qquad
	\Theta^{\mathrm{(p)}}_{\para}(\xp,\rho)=-\frac{1}{\rho}\ln\!\left[\frac{P^{+}_{\od}(\xp,\rho)+P^{-}_{\od}(\xp,\rho)}{2\rme^{-\rho I_{+}(\xp)}}+\frac{P^{+}_{\ev}(\xp,\rho)-\frac{\xp}{|\xp|} P^{-}_{\ev}(\xp,\rho)}{2\rme^{-\rho I_{-}(\xp)}}\right]\!,
\end{equation}
which can be transformed into the results from \cite{HuchtGruenebergSchmidt11} using (\ref{eq:ThetaPerpParaRelation}).

The limit $\rho\to\infty$ is solely covered by $\delta Q_{\od}$, i.e., the original factor $Q_{\od}$ consists of the thermodynamic limit of the bulk contribution and a finite-size part corresponding to the slab geometry oriented in perpendicular direction.
Because of the symmetry $\Theta^{\mathrm{(p)}}_{\para}(\xp,\rho)=\Theta^{\mathrm{(p)}}_{\perp}(\xp,1/\rho)$ the according scaling functions simply read
\begin{equation}
	\Theta^{\mathrm{(p)}}_{\para}(\xp,\rho\to\infty)=-I_{+}(\xp)\qquad\Theta^{\mathrm{(p)}}_{\perp}(\xs,\rho\to0)=-I_{+}(\xs).
\end{equation}
As a consequence, both limits $\rho \to 0$ and $\rho \to \infty$ are invariant under the duality transformation $\xp \mapsto -\xp$.
All contributions in (\ref{eq:Thetap}) are quadratic in $\xp$ except for the sign prefactor of $P_{\ev}^{-}$, where $P_{\ev}^{-}(\xp,\rho\to 0)\to 0$ in the case of thin films.
Thus the topological necessity for the introduction of the four Pfaffians from the dimer approach in (\ref{eq:Z_with_Pfaffians}) breaks the duality symmetry for finite aspect ratios $\rho$.

\section{Scaling functions on the cylinder with open boundary conditions}

If we make a cut along a line on the torus it forms a cylinder, while the boundary conditions are dictated by the two rows of spins along these cut.
The most simple way to do so is to set all the couplings $K^{\perp}_{L}=0$, which results in open boundary conditions.
Our starting point for the according calculation is (\ref{eq:TridiagonalC}), where we set $b=0$.
This is equivalent to making the cut along a line in the parallel direction, and the matrix $\tilde{\mathbf{C}}_{0}$ then is no longer cyclic but only a simple symmetric tridiagonal one, simplifying (\ref{eq:TridiagonalToTM}) to
\begin{equation}
\label{eq:detB0CylinderFinal}
	\det\mathbf{B}_{0} = \det\mathbf{u} \bra{1,0} \tilde{\mathbf{T}}_{L} \tilde{\mathbf{T}}_{L-1}\cdots\tilde{\mathbf{T}}_{2}\tilde{\mathbf{T}}_{1} \ket{1,0},
\end{equation}
as $b_{L}$ vanishes.
As explained in \cite{McCoyWu73} this procedure also reduces the necessity from calculating four determinants to the calculation of merely one, namely $\det\mathbf{A}_{-0}$.
Thus the partition function of the cylinder with homogeneous and isotropic couplings $z=\tanh K$ is given by
\begin{equation}
\label{eq:PartitionFunctionCylinder}
	Z=\left(\frac{2}{1-z^{2}}\right)^{LM} (1-z^2)^{M/2}\sqrt{\det\mathbf{A}_{-0}},
\end{equation}
where the change in the prefactor compared to (\ref{eq:PartitionFunctionTorus}) is due to the missing line of couplings $K^{\perp}_{L}$.
Now we diagonalize $\mathbf{\tilde{T}}_{\ell}$ for all $\ell\neq1$,  while $\mathbf{\tilde{T}}_{1}$ is absorbed into the boundary vector according to $\mathbf{\tilde{T}}_{1}\ket{1,0}=\ket{\mu^{-},1}$, and find
\begin{equation}
\label{eq:detB0Cylinder}
	\det\mathbf{B}_{0}=\left[z(1-z^{2})\right]^{L}\frac{\alpha^{-}(\varphi)\rme^{L\gamma}-\alpha^{+}(\varphi)\rme^{-L\gamma}}{2(1-z^{2})\sinh\gamma},
\end{equation}
with $\alpha^{\pm}(\varphi)=(1-z^{2})\rme^{\pm\gamma}-z^{2}\mu^{-}(\varphi)$.
Again, we can factorize $[z(1-z^2)]^{LM}$ in (\ref{eq:PartitionFunctionCylinder}), which will give non-singular contributions to the bulk and surface terms.
Next we separate (\ref{eq:detB0Cylinder}) into the three parts $Q_{\od}$, $S_{\od}$, and $R_{\od}$, where, in the thermodynamic limit, the first two give the bulk and surface contributions, to obtain
\begin{equation}
	Z=\left(\frac{4z}{1-z^{2}}\right)^{LM/2}\left(1-z^{2}\right)^{M/2}Q_{\od}S_{\od}R_{\od},
\end{equation}
with
\numparts
\begin{eqnarray}
	S_{\od}=\prod_\mystack{0<m<2M}{m\,\mathrm{odd}}\sqrt{
	-\frac{(1-z^{2})\rme^{-\gamma_{m}}+\mu^{-}(\varphi_{m})}{2(1-z^{2})\sinh\gamma_{m}}
	},\\
	R_{\od}=\prod_\mystack{0<m<2M}{m\,\mathrm{odd}}\sqrt{1-
	\frac{(1-z^{2})\rme^{\gamma_{m}}+\mu^{-}(\varphi_{m})}{(1-z^{2})\rme^{-\gamma_{m}}+\mu^{-}(\varphi_{m})}
	\rme^{-2L\gamma_{m}}}.
\end{eqnarray}
\endnumparts
Note that $Q_{\od}$ is the same quantity as in the calculation for the torus, so its residual part $\delta Q_{\od}$ is given by (\ref{eq:IntegralPlus}) and again is solely responsible for the limit $\rho\to\infty$.

\begin{figure}
	\centering
	\hfill%
	\includegraphics[width=0.45\textwidth]{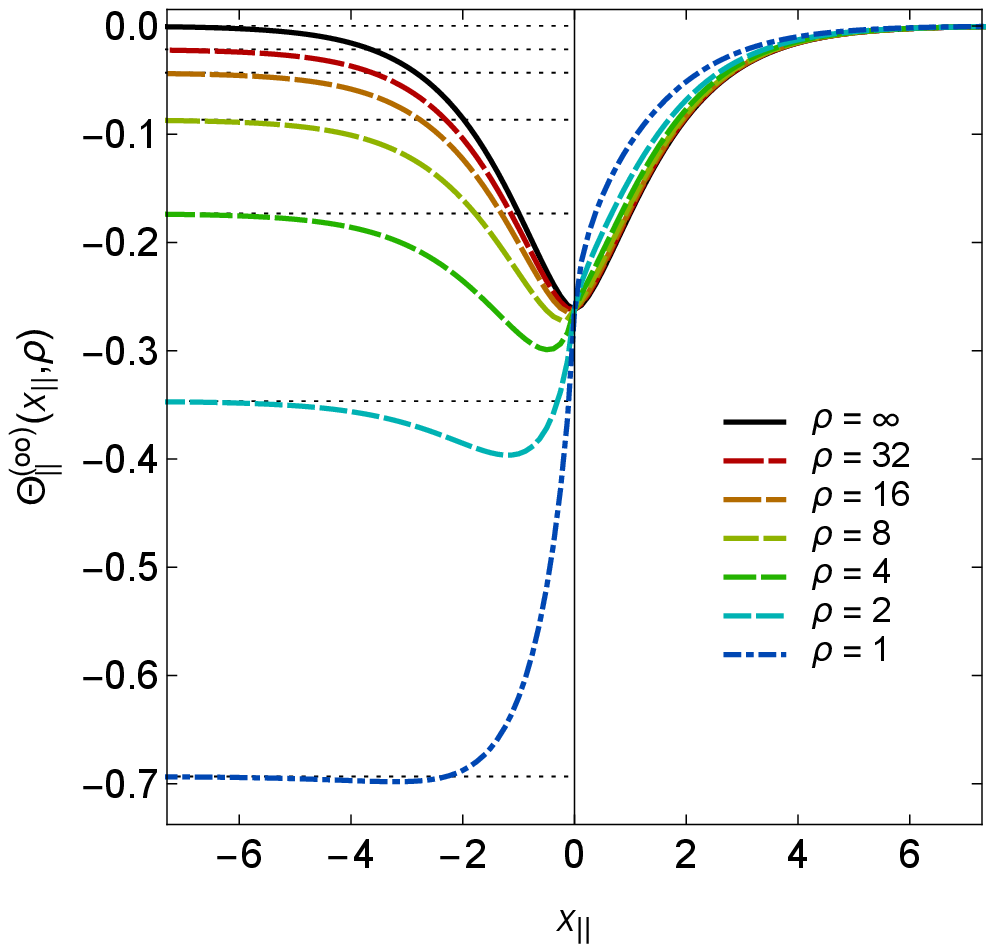}%
	\hfill\hfill%
	\includegraphics[width=0.45\textwidth]{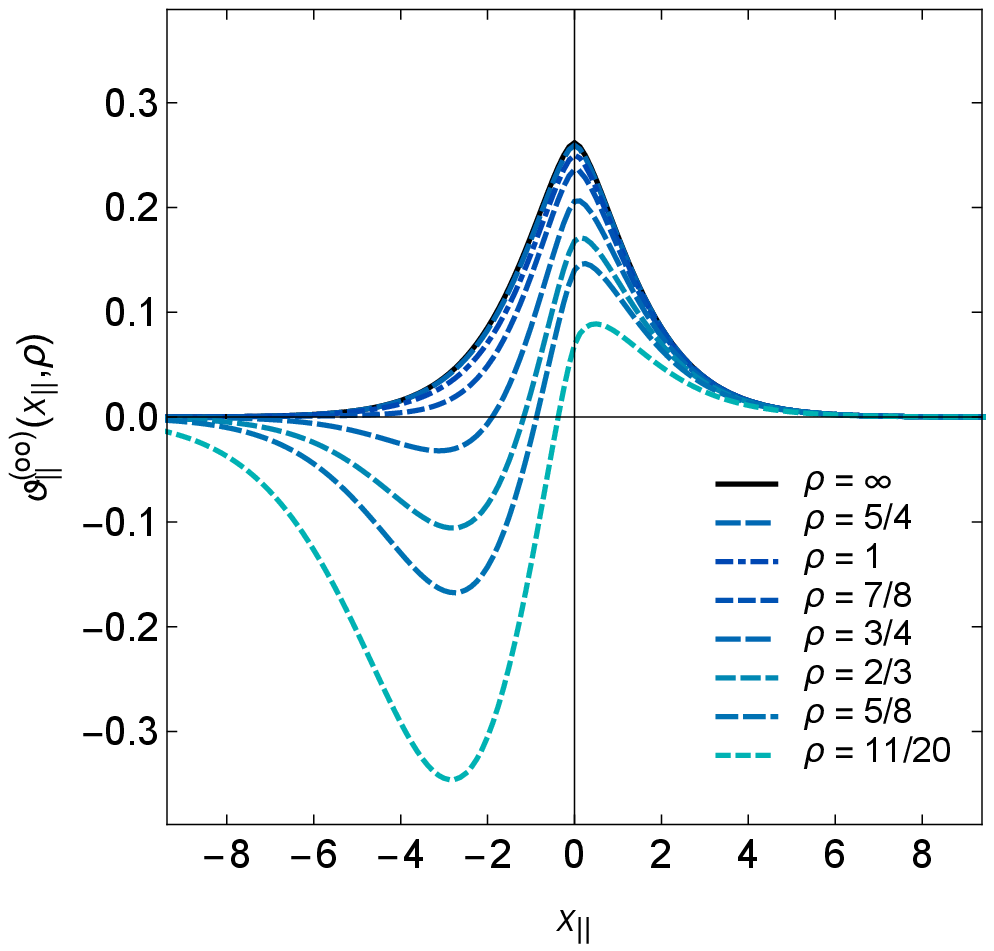}%
	\hfill{}
	\caption{Casimir scaling functions of the cylinder with open boundary conditions for $1 \lesssim \rho \leq \infty$.
	Left: Casimir potential scaling function $\Theta^\mathrm{(oo)}_\para(x_\para,\rho)$. Right: Casimir force scaling function $\vartheta^\mathrm{(oo)}_\para(x_\para,\rho)$.
	}
	\label{fig:ScalingOpenCylinder_para}
\end{figure}

Nevertheless we are interested in the scaling limit and thus we use the regularization technique from the last chapter to obtain the residual part of $S_{\od}$ as well. 
Thus we get an integral along the positive imaginary axis with a singularity at $\omega=\xp$, which can be simplified further as
\numparts
\begin{eqnarray}
	\!\!\! \delta S(\xp)&\equiv&\Re\left\{\frac{1}{2\rmi}\int\limits_{0}^{\rmi\infty}\rmd\omega\,\left[\tan\left(\frac{\pi\omega}{2}\right)-\rmi\right]\ln\left[\frac{1}{2}\left(1+\frac{\xp}{\sqrt{\xp^{2}+\pi^2\omega^{2}}}\right)\right]\right\}\\
	&=&\frac{1}{2}\ln\left[\frac{1}{2}\left(1+\rme^{-|\xp|}\right)\right]-\xp\int\limits_{-\infty}^{\infty}\rmd\omega\,\frac{\ln\left[\frac{1}{2}\left(1+\rme^{-\sqrt{\xp^{2}+4\pi^{2}\omega^{2}}}\right)\right]}{\xp^{2}+4\pi^{2}\omega^{2}}.
\end{eqnarray}
\endnumparts
Analogous to the calculation for $P^{\pm}_{\delta}$ we can formulate $R_{\od}$ as infinite product and find for the scaling limit
\numparts
\begin{eqnarray}
	R_{\od}^{\mathrm{(oo)}}(\xp,\rho)&\equiv&\lim_\mystack{L,M\to\infty}{L/M=\rho}R_{\od}(L,M) \\
	&=&\prod_\mystack{m=-\infty}{m \, \mathrm{odd}}^{\infty}\left(1+\frac{\sqrt{\xp^{2}+\pi^2 m^2}-\xp}{\sqrt{\xp^{2}+\pi^2 m^2}+\xp}\rme^{-2\rho\sqrt{\xp^{2}+\pi^2 m^2}}\right).
\end{eqnarray}
\endnumparts
Finally we find the finite-size scaling function for the Ising model on the cylinder with open boundary conditions to be
\begin{equation}
\label{eq:Thetaoo}
	\Theta_{\para}^{\mathrm{(oo)}}(\xp,\rho)=-\frac{1}{2\rho}\ln\left[\frac{R_{\od}^{\mathrm{(oo)}}(\xp,\rho)}{\rme^{-2[\rho I_{+}(\xp)-\delta S(\xp)]}}\right].
\end{equation}
The critical Casimir force scaling function, calclated with (\ref{eq:Thetathetapara}), then reads
\begin{equation}
\label{eq:thetaoo}
	\vartheta^{\mathrm{(oo)}}_{\para}(\xp,\rho)=I_{+}(\xp)-\!\sum_\mystack{m=-\infty}{m \, \mathrm{odd}}^{\infty}\frac{\sqrt{\xp^{2}+\pi^{2}m^{2}}}{1+\frac{\sqrt{\xp^{2}+\pi^{2}m^{2}}+\xp}{\sqrt{\xp^{2}+\pi^{2}m^{2}}-\xp}\rme^{2\rho\sqrt{\xp^{2}+\pi^{2}m^{2}}}}.
\end{equation}
\begin{figure}
	\centering
	\hfill%
	\includegraphics[width=0.45\textwidth]{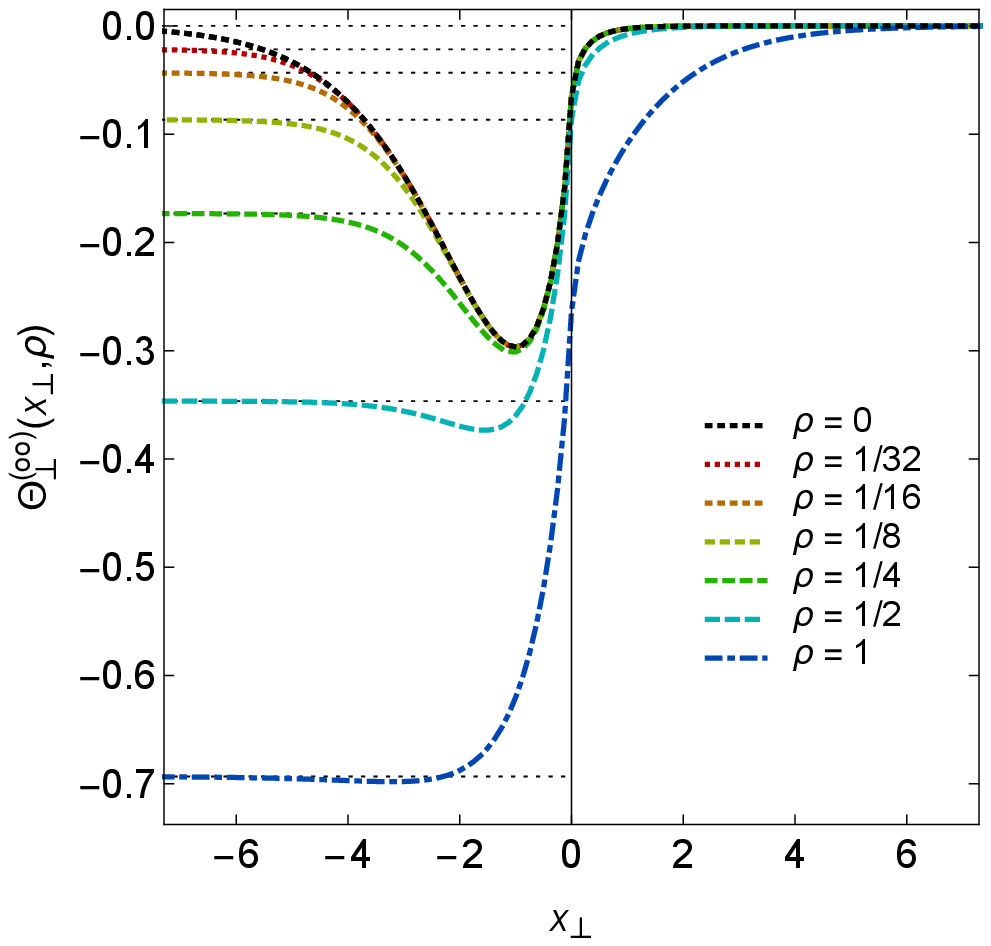}%
	\hfill\hfill%
	\includegraphics[width=0.45\textwidth]{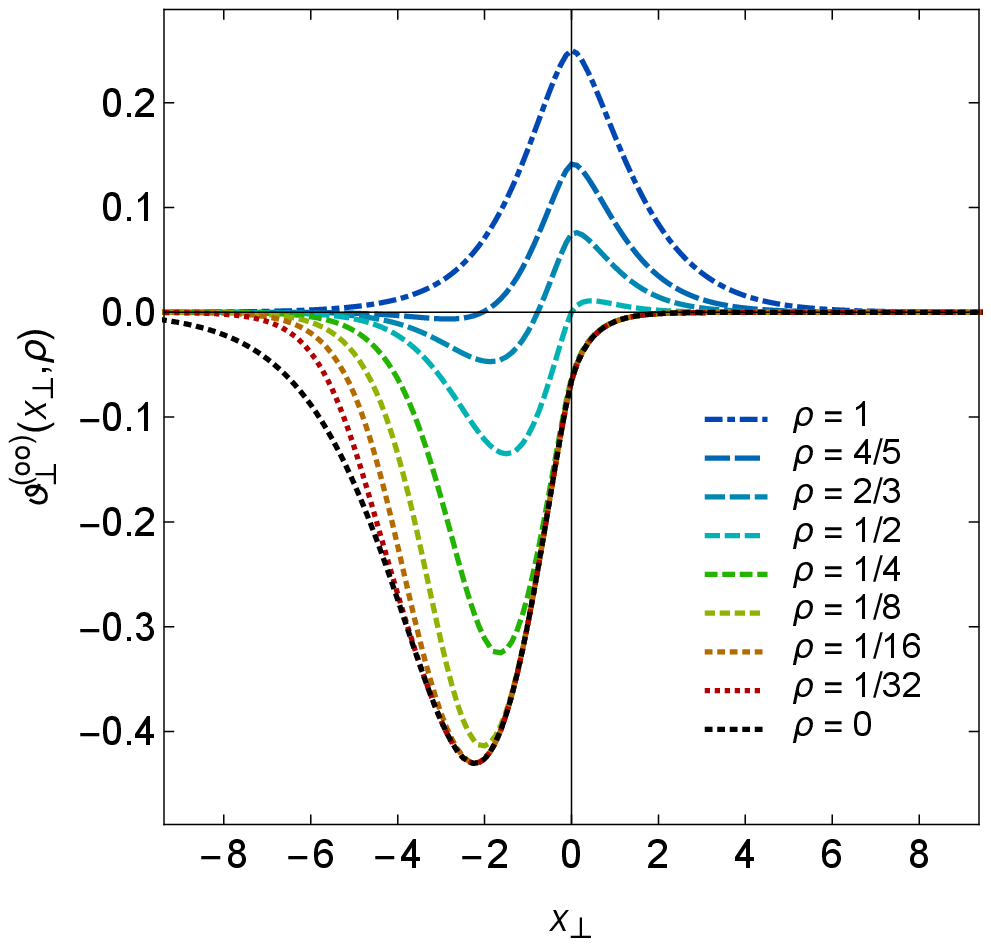}%
	\hfill{}
	\caption{Casimir scaling functions of the cylinder with open boundary conditions for $0 \leq \rho \leq 1$.
	Left: Casimir potential scaling function $\Theta^\mathrm{(oo)}_\perp(x_\perp,\rho)$. Right: Casimir force scaling function $\vartheta^\mathrm{(oo)}_\perp(x_\perp,\rho)$.
	} \label{fig:ScalingOpenCylinder_perp}
\end{figure}

It is worth mentioning, that for the torus as well as for the cylinder with open boundary conditions for $\xs\to -\infty$ the scaling functions (\ref{eq:Thetap}) and (\ref{eq:Thetaoo}) both simplify to
\begin{equation}
	\Theta_{\para}^{\mathrm{(p)/(oo)}}(\xs\to-\infty,\rho)=-\frac{\ln 2}{\rho},
\end{equation}
which is a direct consequence of the broken symmetry in the ordered phase \cite{HuchtGruenebergSchmidt11}.

For $\rho\to0$ this scaling function is identical to the well-known solution for the thin film geometry \cite{EvansStecki94}; to see this, we change to the scaling variable in perpendicular direction $\xs$ and use the Euler-Maclaurin formula to reformulate $\ln R_{\od}^{\mathrm{(oo)}}(\xs,\rho)$ as integral.
With the substitution $\omega=(2m-1)\pi\rho$ it becomes independent of $\rho$, while the first correction is already linear in the aspect ratio, thus vanishing in the aspired limit.
It is easy to see that the other two terms in $\xs$ also do vanish for $\rho\to0$, leading to 
\begin{equation}
	\Theta_\perp^{\mathrm{(oo)}}(\xs,\rho\to0)=-\frac{1}{2\pi}\int\limits_{0}^{\infty}\rmd\omega\,\ln\left[1+\frac{\sqrt{\xs^{2}+\omega^{2}}-\xs}{\sqrt{\xs^{2}+\omega^{2}}+\xs}\rme^{-2\sqrt{\xs^{2}+\omega^{2}}}\right]
\end{equation}
and the according Casimir force scaling function
\begin{equation}
\label{eq:thetaoofilm}
	\vartheta_\perp^{\mathrm{(oo)}}(\xs,\rho\to0)=-\frac{1}{\pi}\int\limits_{0}^{\infty}\rmd\omega\,\frac{\sqrt{\xs^{2}+\omega^{2}}}{1+\frac{\sqrt{\xs^{2}+\omega^{2}}-\xs}{\sqrt{\xs^{2}+\omega^{2}}+\xs}\rme^{2\sqrt{\xs^{2}+\omega^{2}}}},
\end{equation}
which gives indeed the correct limit \cite{EvansStecki94}, as shown in figure~\ref{fig:ScalingOpenCylinder_perp} as dotted line.
Note that the sum in (\ref{eq:thetaoo}) is the generalization of the integral in (\ref{eq:thetaoofilm}) for arbitrary aspect ratio $\rho$.

If we set $x=0$ we come to the regime of conformal field theory.
First we notice that $I_{+}(0)=-\frac{\pi}{12}$ and that $\delta S(0)=0$. The product $R_{\od}^{\mathrm{(oo)}}(0,\rho)$ simplifies to
\begin{equation}
	R_{\od}^{\mathrm{(oo)}}(\xp=0,\rho)=\prod_{m=-\infty \atop m \, \mathrm{odd}}^{\infty}\left(1+\rme^{2\pi\rho|m|}\right)=2\frac{\left(-1,\rme^{-2\pi\rho}\right)_{\infty}}{\left(-1,\rme^{-4\pi\rho}\right)_{\infty}},
\end{equation}
where $(a,q)_{\infty}$ is the $q$-Pochhammer symbol, and we get the scaling function
\begin{equation}
	\Theta_{\para}^{\mathrm{(oo)}}(\xp=0,\rho)=-\frac{\pi}{12}-\frac{1}{\rho}\ln\frac{\left(-1,\rme^{-2\pi\rho}\right)_{\infty}}{\left(-1,\rme^{-4\pi\rho}\right)_{\infty}}.
\end{equation}
As there is no explicitly preferred direction in the conformal field theory, its scaling functions are those of the volume, see (\ref{eq:FresVolume}), and thus we have to change to
\begin{equation}
	\Theta(x=0,\rho)=-\frac{\pi}{12}\rho-\ln\frac{\left(-1,\rme^{-2\pi\rho}\right)_{\infty}}{\left(-1,\rme^{-4\pi\rho}\right)_{\infty}}.
\end{equation}
This can be recasted to the conformal field theory result
\begin{equation}
	\Theta(x=0,\rho)=-\frac{1}{2}\ln\frac{\theta_{3}(\rme^{-2\pi\rho})}{\eta(2\rmi\rho)},
\end{equation}
where $\theta_{3}(q)=\theta_{3}(z=0|\rme^{\rmi\pi\tau})$ is a Jacobi theta function and $\eta(\tau)$ is the Dedekind eta function \cite{Philippe97}, as both can be expressed in terms of $q$-Pochhammer symbols after some algebra.
Note that the conformal field theory result is usually given for the annulus, but the free energy of the cylinder $F_{\mathrm{cyl}}$ and the free energy of the annulus $F_{\mathrm{ann}}$ at criticality are connected via the relation $F_{\mathrm{ann}}=F_{\mathrm{cyl}}+\frac{\pi}{12}\rho$, see \cite{BimonteEmigKardar13}.

\section{Conclusion}

We presented a systematic calculation of Casimir force scaling functions and the Casimir potential scaling functions for the homogeneous and isotropic Ising model on the torus and on the cylinder with open boundary conditions.
Therefore we started with the dimer representation of the according partition function and reduced the necessary calculations of Pfaffians of $4LM\times4LM$ matrices to a $2\times2$ transfer matrix representation.
We introduced a way to regularize the occurring integrals representing the bulk and the surface parts to get the finite-size contributions to the free energy in the scaling limit.
Interestingly, for both Casimir potential scaling functions the cases $\rho\to\infty$ are covered by the regularized bulk contributions, while for the cylinder with finite aspect ratio $\rho$ an additional surface contribution arises.
For the other geometrical limiting case $\rho\to0$, our results lead to the correct limits of thin films for both cases and agree with the predictions of conformal field theory at criticality $x=0$.

As done in \cite{EvansStecki94} for the infinite strip, it should be possible to expand this calculation to other boundary conditions, too, like $(++)$ or $(+-)$ boundaries instead of open boundaries to obtain aspect ratio depending scaling functions.
Those should coincide, at $x=0$, with the according results from conformal field theory \cite{BimonteEmigKardar13}.
Finally we mention that the case of open boundary conditions in both directions has recently been studied in \cite{Hucht16a,Hucht16b}.

\ack{
This work was supported by the Deutsche Forschungsgemeinschaft through Grant No.~HU 2303/1-1.
}

\section*{References}

\bibliographystyle{unsrt}
\bibliography{Physik}

\begin{thebibliography}{10}

\bibitem{Onsager44}
L.~Onsager.
\newblock Crystal statistics. {I.} {A} two-dimensional model with an
  order-disorder transition.
\newblock {\em Phys. Rev.}, 65:117, 1944.

\bibitem{Widom65}
B.~Widom.
\newblock Equation of state in the neighborhood of the critical point.
\newblock {\em J.~Chem. Phys.}, 43:3898, 1965.

\bibitem{Kadanoff66}
L.~P. Kadanoff.
\newblock Scaling laws for {Ising} models near {$T_c$}.
\newblock {\em Physics}, 2:263, 1966.

\bibitem{Polyakov70}
A.~M. Polyakov.
\newblock Conformal symmetry of critical fluctuations.
\newblock {\em JETP Lett.}, 12(12):381, 1970.

\bibitem{Cardy84}
J.~Cardy.
\newblock Conformal invariance and universality in finite-size scaling.
\newblock {\em J.~Phys A: Math. Gen.}, 17:L385, 1984.

\bibitem{CardyPeschel88}
John Cardy and Ingo Peschel.
\newblock Finite-size dependence of the free energy in two-dimensional critical
  systems.
\newblock {\em Nucl. Phys.~B}, 300:377, 1988.

\bibitem{Kasteleyn63}
P.~W. Kasteleyn.
\newblock Dimer statistics and phase transitions.
\newblock {\em J.~Math. Phys.}, 4:287, 1963.

\bibitem{McCoyWu73}
B.~M. McCoy and T.~T. Wu.
\newblock {\em The Two-Dimensional {Ising} Model}.
\newblock Harvard University Press, Cambridge, 1973.

\bibitem{Casimir48}
H.~B.~G. Casimir.
\newblock On the attraction between two perfectly conducting plates.
\newblock {\em Proc. K. Ned. Akad. Wet.}, 51:793, 1948.

\bibitem{FisherdeGennes78}
M.~E. Fisher and P.-G. de~Gennes.
\newblock Ph\'enom\`enes aux parois dans un m\'elange binaire critique.
\newblock {\em C. R. Acad. Sci. Paris, Ser. B}, 287:207, 1978.

\bibitem{GarciaChan99}
R.~Garcia and M.~H.~W. Chan.
\newblock Critical fluctuation-induced thinning of $^4${He} films near the
  superfluid transition.
\newblock {\em Phys. Rev. Lett.}, 83:1187, 1999.

\bibitem{GarciaChan02}
R.~Garcia and M.~H.~W. Chan.
\newblock Critical {Casimir} effect near the $^3${He}-$^4${He} tricritical
  point.
\newblock {\em Phys. Rev. Lett.}, 88:086101, 2002.

\bibitem{FukutoYanoPershan05}
M.~Fukuto, Y.~F. Yano, and P.~S. Pershan.
\newblock Critical {Casimir} effect in three-dimensional {Ising} systems:
  Measurements on binary wetting films.
\newblock {\em Phys. Rev. Lett.}, 94:135702, 2005.

\bibitem{HertleinHeldenGambassiDietrichBechinger08}
C.~Hertlein, L.~Helden, A.~Gambassi, S.~Dietrich, and C.~Bechinger.
\newblock Direct measurement of critical {Casimir} forces.
\newblock {\em Nature}, 451:172, 2008.

\bibitem{SoykaZvyaHertHeldBech08}
Florian Soyka, Olga Zvyagolskaya, Christopher Hertlein, Laurent Helden, and
  Clemens Bechinger.
\newblock Critical {Casimir} forces in colloidal suspensions on chemically
  patterned surfaces.
\newblock {\em Phys. Rev. Lett.}, 101:208301, Nov 2008.

\bibitem{BonnOtwiSacaGuoWegSchall09}
Daniel Bonn, Jakub Otwinowski, Stefano Sacanna, Hua Guo, Gerard Wegdam, and
  Peter Schall.
\newblock Direct observation of colloidal aggregation by critical {Casimir}
  forces.
\newblock {\em Phys. Rev. Lett.}, 103:156101, Oct 2009.

\bibitem{ZAB11}
O.~Zvyagolskaya, A.~J. Archer, and C.~Bechinger.
\newblock Criticality and phase separation in a two-dimensional binary
  colloidal fluid induced by the solvent critical behavior.
\newblock {\em EPL (Europhysics Letters)}, 96(2):28005, 2011.

\bibitem{DVNBS13}
Minh~Triet Dang, Ana~Vila Verde, Van~Duc Nguyen, Peter~G. Bolhuis, and Peter
  Schall.
\newblock Temperature-sensitive colloidal phase behavior induced by critical
  {Casimir} forces.
\newblock {\em The Journal of Chemical Physics}, 139(9):094903, 2013.

\bibitem{Hucht07a}
Alfred Hucht.
\newblock Thermodynamic {Casimir} effect in $^{4}${He} films near
  {$T_\lambda$}: {Monte} {Carlo} results.
\newblock {\em Phys. Rev. Lett.}, 99(18):185301, Nov 2007.

\bibitem{VasilyevGambassiMaciolekDietrich09}
O.~Vasilyev, A.~Gambassi, A.~Macio{\l}ek, and S.~Dietrich.
\newblock Universal scaling functions of critical {Casimir} forces obtained by
  {Monte} {Carlo} simulations.
\newblock {\em Phys. Rev.~E}, 79(4):041142, 2009.

\bibitem{Hasenbusch13}
Martin Hasenbusch.
\newblock Thermodynamic {Casimir} forces between a sphere and a plate: {Monte}
  {Carlo} simulation of a spin model.
\newblock {\em Phys. Rev. E}, 87:022130, Feb 2013.

\bibitem{Vasilyev14}
O.~A. Vasilyev.
\newblock Critical {Casimir} interactions between spherical particles in the
  presence of bulk ordering fields.
\newblock {\em Phys. Rev. E}, 90:012138, Jul 2014.

\bibitem{HobrechtHucht14}
Hendrik Hobrecht and Alfred Hucht.
\newblock Direct simulation of critical {C}asimir forces.
\newblock {\em EPL}, 106(5):56005, Jun 2014.
\newblock {arXiv:1405.4088}.

\bibitem{HobrechtHucht15a}
Hendrik Hobrecht and Alfred Hucht.
\newblock Many-body critical {C}asimir interactions in colloidal suspensions.
\newblock {\em Phys. Rev.~E}, 92:042315, 2015.

\bibitem{BurkhardtEisenriegler95}
Theodore~W. Burkhardt and Erich Eisenriegler.
\newblock {Casimir} interaction of spheres in a fluid at the critical point.
\newblock {\em Phys. Rev. Lett.}, 74:3189--3192, Apr 1995.

\bibitem{EisenrieglerRitschel95}
E.~Eisenriegler and U.~Ritschel.
\newblock {Casimir} forces between spherical particles in a critical fluid and
  conformal invariance.
\newblock {\em Phys. Rev. B}, 51:13717--13734, May 1995.

\bibitem{BimonteEmigKardar13}
G.~Bimonte, T.~Emig, and M.~Kardar.
\newblock Conformal field theory of critical {Casimir} interactions in {2D}.
\newblock {\em EPL (Europhysics Letters)}, 104(2):21001, 2013.

\bibitem{LuWu01}
Wentao~T. Lu and F.~Y. Wu.
\newblock {Ising} model on nonorientable surfaces: {Exact} solution for the
  {M{\"o}bius} strip and the {Klein} bottle.
\newblock {\em Phys. Rev.~E}, 63:026107, 2001.

\bibitem{Philippe97}
Philippe~Di Francesco, P.~Mathieu, and D.~Senechal.
\newblock {\em Conformal Field Theory}.
\newblock Graduate Texts in Contemporary Physics. Springer, 1997.

\bibitem{FerdinandFisher69}
A.~E. Ferdinand and M.~E. Fisher.
\newblock Bounded and inhomogeneous {Ising} models. {I.} {Specific}-heat
  anomaly of a finite lattice.
\newblock {\em Phys. Rev.}, 185:832, 1969.
\newblock There is a typo in Eq. (3.36), the term $\xi S_{1}(n)\tau^{2}/2$ is
  missing.

\bibitem{HuchtGruenebergSchmidt11}
Alfred Hucht, Daniel Gr\"uneberg, and Felix~M. Schmidt.
\newblock Aspect-ratio dependence of thermodynamic {Casimir} forces.
\newblock {\em Phys. Rev.~E}, 83:051101, Mar 2011.
\newblock {arXiv:1012.4399}. {There} is a '$-$' missing in the second term of
  Eq.~(47b), and '$+$' and '$-$' are interchanged in the sentence before.

\bibitem{Kaufman49}
B.~Kaufman.
\newblock Crystal statistics. {II}. {Partition} function evaluated by spinor
  analysis.
\newblock {\em Phys. Rev.}, 76(8):1232--1243, Oct 1949.

\bibitem{EvansStecki94}
R.~Evans and J.~Stecki.
\newblock Solvation force in two-dimensional {Ising} strips.
\newblock {\em Phys. Rev. B}, 49:8842--8851, Apr 1994.

\bibitem{Dohm09}
V.~Dohm.
\newblock Critical {Casimir} force in slab geometry with finite aspect ratio:
  Analytic calculation above and below {$T_c$}.
\newblock {\em EPL}, 86(2):20001, 2009.

\bibitem{GradshteynRyzhik80}
I.~S. Gradshteyn and I.~M. Ryzhik.
\newblock {\em Table of integrals, series, and products; corrected and enlarged
  edition}.
\newblock Academic Press, London, 1980.

\bibitem{Hucht16a}
Alfred Hucht.
\newblock The square lattice {Ising} model on the rectangle {I}: {Finite}
  systems.
\newblock {\em J.~Phys A: Math. Theor.}, 2016.
\newblock submitted, {arXiv:1609.01963}.

\bibitem{Hucht16b}
Alfred Hucht.
\newblock The square lattice {Ising} model on the rectangle {II}: {Finite}-size
  scaling limit.
\newblock in preparation.

\end{thebibliography}

\end{document}